# The folding mechanics of a knotted protein

Stefan Wallin, Konstantin B Zeldovich, Eugene I Shakhnovich.

Department of Chemistry & Chemical Biology, Harvard University, 12 Oxford Street, Cambridge MA 02138, USA.

Corresponding author: Eugene I Shakhnovich.

Classification: BIOLOGICAL (major) and BIOPHYSICS (minor).

Number of text pages: 14

Number of Figures: 7.

Number of words in abstract: 216.

Number of characters (total): 46,985.

**Abstract**

An increasing number of proteins are being discovered with a remarkable and somewhat surprising feature, a knot in their native structures. How the polypeptide chain is able to "knot" itself during the folding process to form these highly intricate protein topologies is not known and represents a novel challenge to the protein folding community. To address this question we perform a computational study on the 160-amino acid homodimeric protein YibK which, like other proteins in the SpoU family of MTases, contains a deep trefoil knot in its C-terminal region. In this study, we use a coarse-grained $C_\alpha$-chain representation and Langevin dynamics to study folding kinetics. By testing different ways of adding interactions to a native-centric potential, we find that *specific*, attractive nonnative interactions are critical for knot formation. In the absence these interactions, i.e. in an energetics driven entirely by native interactions, knot formation is exceedingly unlikely. Focusing on an optimal potential, we find, in concert with recent experimental data on YibK, two parallel folding pathways which we attribute to an early and a late formation of the trefoil knot, respectively. For both pathways, knot formation occurs before dimerization. A bioinformatics analysis of the SpoU family of proteins reveals further that the critical nonnative interactions may originate from evolutionary conserved hydrophobic segments around the knotted region.

**Introduction**

Are there knots in proteins? This question was first posed in 1994 by Mansfield (1). At the time, not more than about 400 protein structures were available for analysis and only two potential cases of very shallow knots could be identified. For such borderline cases, classifying proteins as knotted is nontrivial as knots are only rigorously defined for closed loops (how should the ends be connected?) (2). Since then, the issue has been revisited several times and additional examples of more deeply knotted proteins have surfaced (3-6), and we now undoubtedly have to conclude that knots do indeed occur in proteins. In the most recent review (6), 273 proteins with some type of knotted conformation were found in the PDB. Still, knots in proteins are relatively rare and most examples of knots are simple trefoils, i.e. they have 3 projected crossings. The known set of knotted proteins is dominated by two classes (6): bacterial and viral rRNA methyltransferases, which make up the α/β-knot superfamily (7), and various isozymes of carbonic anhydrase. In addition, a few other insular examples of knots in proteins have been found, one of which is ubiquitin hydrolase (UCH-L3), the most complicated knot in a human protein found to date, with five projected crossings (6).

Here we address the question: What are the mechanics of knot formation in proteins during the folding process? To our knowledge, this issue has not been previously addressed. One likely reason for this is the difficulty in designing suitable experiments that can directly probe knot formation as it is a subtle conformational transition. We thus turn to computer simulations in which, at least in principle, every conformation from the unfolded to the native state is available for analysis. To monitor knot formation in our simulations we make use of KMT-reduction (8), a test procedure where atoms are sequentially removed if it does not alter the "topology" of the chain. If the test procedure reduces the chain to its first and last atoms the chain is considered the unknot (atom j is removed if the triangle defined by atoms $j-1$, $j$, and $j+1$ is not intersected by any other bond vector in the chain). Despite the lack of mathematical rigor, we find the KMT reduction method reliable and very useful as an operational definition of knot occurrences in simulations.

We focus here on the homodimeric protein YibK (PDB id: 1J85) which belongs to the SpoU family of MTases and is well-suited for a computational study. It contains 160 amino acids and has a deep trefoil knot at approximately 40 residues from its C-terminal (9), as depicted in Fig. 1. In recent works, Mallam and Jackson extensively characterized (10-12) the kinetics and thermodynamics of this protein, although details of knot formation could not be addressed. It was shown that YibK folds reversibly *in vitro* indicating that chaperon activity is not a prerequisite for folding, and it was revealed that YibK has a multi-phasic folding kinetics, with the slowest phase corresponding to dimerization. Moreover, the experimental data suggested that monomeric forms of the protein dominate at low pH and low protein concentration. For this, and for computational reasons, we focus here mainly on the monomer folding of YibK. We start by straightforwardly applying a well-studied and commonly used Gō-type model (13) to the folding of YibK. We find that for YibK, this standard Gō-type model, in which only native interactions are treated as attractive, fails at producing even a plausible folding scenario as knot formation does not occur. Only by modifying the energetics of the system through the addition of certain *specific* nonnative, attractive interactions, can we achieve a successful folding behavior, highlighting the importance of these types of forces during the folding of this protein. In addition to our computer simulations, we perform a bioinformatics analysis of the SpoU family of MTases in an attempt to find evidence for conserved sequence characteristics that relates to knot formation. As knots in proteins are usually conserved within protein families (6), we expect most members of the SpoU family to contain similar types of trefoil knots as in the YibK protein. Indeed, several recently determined structures within this family have been found to have deep trefoil knots (14-17). Through an analysis where we compare conserved sequence properties of the SpoU family with those obtained from structurally similar but unknotted protein structures, we find evidence for some unusual sequence properties in the SpoU family close to the trefoil knot, which may be linked to critical nonnative interactions during knot formation.

**Results and Discussion**
Our initial goal is to investigate the role of native interactions for the folding dynamics of YibK. We therefore apply a well-studied coarse-grained Gō-type $C_\alpha$ model (13, 18-22) to the YibK monomer, in which only interactions present in the native structure are attractive, and explore the kinetic behavior using Langevin dynamics (see Models and Methods). This type of "native-centric" model has been used widely in the past several years to study folding kinetics and thermodynamics because of computational limitations on more realistic sequence-based all-atom protein force fields (23-25). As nonnative interactions are ignored, the native-centric approach to protein modeling usually guarantees folding to a proteins' native state, and some justification for its use exists (26). In particular, it has been found useful when tested against certain experimental data (19, 27, 28).

We find, however, that an entirely native-centric approach can not explain the folding behavior of the YibK protein. Nonetheless, it is illuminating to explore the kinetic behavior obtained for the coarse-grained $C_\alpha$ model. Figure 2*a* shows the time evolution of the fraction of native contacts, $Q$, in a typical simulation for the YibK monomer started from a random configuration, at $T = 0.975$, which is slightly below the melting temperature, $T_m$. A reversible transition to a state characterized by $Q \approx 0.7$-$0.8$ occurs. This state is compact and quite native-like in terms of the cRMSD similarity measure (data not shown), but the chain remains unknotted and is thus unable to reach its native state. It should be pointed out that, by construction, the energy-minimum state of this model is very close to the native structure, which, of course, in the case of YibK includes a trefoil knot. It could therefore be suggested from Fig. 2*a* that folding does not occur because the native state is kinetically inaccessible from the unfolded state. This is not the case, however. To show this, we perform several additional simulations started from the native structure at a slightly elevated temperature $T = 1.04 > T_m$. In all cases, unfolding is readily observed. A typical unfolding trajectory can be seen in Fig. 2*b*. Hence, we can conclude that the unfolded and native states are kinetically connected in our model, but in a dynamic process dictated by native interactions alone the probability of reaching the native state is exceedingly low. Clearly, interactions critical to the folding of YibK are absent from this standard Gō-model.

In order to achieve a successful folding behavior for our knotted protein, it is necessary to include attractive, nonnative interactions in our model. Moreover, we find that these forces must be included in a rather specific manner. We tested a number of different ways to add interactions to the Gō-forces of the native-centric $C_\alpha$ model (see Supporting Information). Of these, only one exhibits a reliable folding behavior, consistent with the reversible folding behavior observed for YibK (10, 11). This improved model is characterized by, in addition to the original Gō-forces, nonnative attractions between two separate regions of the protein chain, the chain "end" from the native trefoil knot to the C-terminus (positions 122-147) and a region roughly in the middle of the knot (positions 86-93). These regions are highlighted in the native structure in Fig. 1. Although our results are robust with respect to small changes in the position and size of these two regions, different ways of including nonnative attractions yielded significantly less reliable folding behavior. For example, the "global," *nonspecific* prescription of including nonnative interactions used in a recent study (29) of the present $C_\alpha$ model, which generally increased folding rates at moderate nonnative interaction strengths, does not provide any improvement of the knot formation ability vis-à-vis the original Gō-model. In the following, we will therefore focus on this improved "nonnative" $C_\alpha$ model, which reliably and reproducibly folds YibK into its native state.

Figure 3*a* shows an example of a folding trajectory obtained using the nonnative $C_\alpha$ model,

at $T = 0.96$. In sharp contrast to the original Gō-model, the trefoil knot is readily formed and the native state can be reached. Interestingly, despite the very different folding behavior between the original $C_\alpha$ model and the nonnative model, the energetic contribution of the nonnative interactions is very small compared to the native interactions, as can be seen in Fig. 3*b*. The effect of the nonnative interactions on the overall thermodynamic properties is therefore likely very limited. Instead, the improved "success rate" for folding must originate from favorable changes in the folding dynamics leading to an increased kinetic accessibility of the native state. How this occurs can be understood in the following way: the native trefoil knot in YibK is located approximately 40 residues from the C-terminus and 70 residues from the N-terminus. Assuming therefore that the knot is formed at the C-terminal region, the C-terminus must enter a loop formed by a preceding chain segment (which may or may not be in its native conformation) in order to form the native trefoil knot. In our model, this "loop entrance" by the C-terminus is effectively mediated by the added attractive nonnative interactions. The effect of these transient nonnative interactions during knot formation can be seen in greater detail in Fig. 4, which shows snapshots of a folding event including the knot formation. (A movie of this folding trajectory can be seen at http://www-shakh.harvard.edu/movies/yibk.html.)

In order to examine more quantitatively the folding process of the YibK monomer, we construct an ensemble of structures in the following way: From 1000 folding simulations we select, for each successful folding trajectory, the first occurring conformation which contains a knot. Note that a conformation is selected only if the chain proceeds to fold directly without any further unknotting/knotting of the chain, so that only one conformation is selected per folding event. Hence, this "knot transition state ensemble" represents structures at the transition between unknotted to knotted states during folding. Figure 5 shows the probability distribution of $Q^*$, where $Q^*$ is the fraction of native contacts for this knot transition state. The bimodal shape of the distribution shows that folding occurs through one of two parallel "pathways," distinguished by a late ($Q^* \approx 0.8$) or an early ($Q^* \approx 0.2$) formation of the trefoil knot. In fact, the folding trajectories in Figs. 3 and 4, respectively, are representative examples of these two folding pathways. As can be seen from Fig. 3*a*, the "late-knot" pathway is associated with an (unknotted) intermediate state during folding. This intermediate is quite native-like with approximately 70% of native contacts formed. In the $Q^* = 0.2$ folding pathway, the chain proceeds quickly to the native state once the trefoil knot has been formed (data not shown). Hence, the corresponding intermediate, which contains a knot but otherwise closely resemble conformations in the unfolded state, is short-lived in our model.

Another interesting question in terms of the knot transition state is the size and location of these initial knots. The first and last amino acids of a knot can be determined by repeatedly applying the KMT-reduction scheme, noting the first instance at which no knot is detected as amino acids are sequentially removed from one chain end. In the native structure of YibK, for example, this procedure leads to a knot at positions 75-121. Figure 5 *inset* shows the probability distributions of the first and last residues of the knots in our set of transition conformations. These distributions show that, in most case, these initial knots stretch approximately positions 75 to 147, meaning that the knots are usually formed by the threading of the C-terminus. In some of the cases, however, the last residue of a knot is found at around position 121, close to the native position. Visual inspection of these structures reveals that in these cases the knot is formed by threading not the C-terminus directly, but a preceding part of the chain folded in a hairpin-like manner. In the majority of the cases, however, the knot is formed by the threading of the C-terminus followed by a tightening from the C-terminal end to form the fully native trefoil knot stretching positions 75 to 121.

Finally, we address the dimerization of YibK. To this end, we perform 100 folding simulations with 2 interacting chains in a cube with periodic boundary conditions, using our nonnative $C_\alpha$ model at $T = 0.96$. Successful folding to the native dimer occurs in 45 of these trajectories. In the remaining trajectories, binding occurs with one, or both, chains in the unknotted intermediate state resulting in a trapped "misfolded" dimerized state. None of these proceed to form the correct dimer form. In all cases with successful folding and binding, knot formation occurs before the dimerization step, as illustrated in Fig. 6. A likely reason for the observed misfolding events is the high chain density used in our 2-chain simulations, which was chosen for computational efficiency. At a lower chain density, each monomer would have a greater probability of forming its knot before contacting another chain and thus dimerizing successfully. Hence, based on the consistent behavior of the successful dimerization events, we find that our model suggests that knot formation occurs before dimerization, i.e. in the first phase of the folding process of the YibK protein.

The obtained folding behavior for our model can be compared to experimental data on the folding kinetics of YibK obtained by Mallam and Jackson (10-12). They found their data consistent with two parallel pathways, with structurally different intermediates $I_1$ and $I_2$, leading to a single monomer state which subsequently dimerizes in a slow, rate-limiting step to the native dimer (see

Fig. 7 in Ref. (11). This folding scenario fits remarkably well with our simulations results and allows us to discuss various aspects of the proposed folding process of YibK as it relates to knot formation. Based on our simulation results, we identify the two parallel pathways with an "early" and a "late" formation of the trefoil knot. Moreover, these two pathways are associated with a poorly structured (and knotted) and native-like (and unknotted) intermediate states, respectively. Also, as mentioned above, we find that knot formation occurs before the dimerization step, regardless of which folding pathway is taken in the monomer folding phase.

In Ref. (11), it was concluded that the parallel pathways in the monomer folding originate from a heterogeneity in the proline isomer population in the denatured state; 1 of the 10 prolines in YibK is natively in its intrinsically unfavorable *cis* conformation. Our $C_\alpha$ coarse-grained chain model is, of course, unable to address folding in such detail. It could therefore be suggested that in the absence of such details in our model, we should observe only one of the folding pathways, namely $I_2$, as it corresponds to the folding pathway initiated from the unfolded state where all proline isomers are in native-like conformations. However, it should be pointed out that the non-native prolyl-peptidyl species in the denatured state cause folding through a different, and structurally distinct, folding intermediate $I_1$, and not by a slower parallel reaction to the same species which has been observed for a number of other proteins (11). It is therefore possible that intrinsic parallel pathways exist for the YibK monomer, which we see here despite the coarse-grained nature of our model.

## Sequence characteristics of the SpoU family of proteins

Our chain modeling results suggest that certain specific nonnative interactions are critical to the knot formation of YibK. An important question is then, of course, if signatures of such interactions can be discerned from the amino acid sequence properties of YibK. A natural candidate for these nonnative attractive forces is the hydrophobic effect, as it is strongly attractive and one of the main driving forces for folding. Before turning to this question, however, it must be pointed out that our simulations results are applicable to any protein with a similar native structure regardless of the particular amino acid sequence, as for any native-centric-type model. Hence, the validity of the proposed nonnative interactions is more accurately discussed in terms of the sequence properties of the SpoU family of proteins rather than the YibK sequence alone, as we expect most of these proteins to have native topologies similar to YibK (including the deep trefoil knot) (30).

In our quest for sequence signals responsible for knot formation, we apply the following logic. It is known that sequence properties can partially be attributed to local structural propensities, i.e. a significant correlation between local structure and sequence properties exisits (31). Knot formation, on the other hand, is a global event as it involves specific non-local (along the sequence) amino acid pairs being brought into close proximity. Therefore, we seek sequence patterns located around the knotted region which can not be attributed to local structural propensities. To carry out this program, we construct first a multiple sequence alignment (32) of the approximately 1500 sequences available (33) to date for the SpoU family of proteins. The average hydrophobicity profile of this SpoU sequence alignment is shown in Fig. 7*a* (*red curve*). Second, to find the role of local interactions in this profile, we construct a reference alignment (*green curve*) based on the *local* structure of the YibK protein, as described in Models and Methods. At each position, this reference alignment represents the typical amino acid composition for small fragments with roughly the same local structure as in YibK. Now, we note in particular two regions which are strongly hydrophobic in the SpoU alignment. These are located in the middle of the native trefoil knot ($\beta_5$) and in the C-terminal end region (positions 139-142 of $\alpha_5$). It is not unreasonable to expect that attractions between these two regions might have an effect similar to the nonnative attractive forces introduced in our $C_\alpha$ model. Further, we note that these two "hydrophobic signals" deviate clearly from the reference alignment, indicating that they can not be attributed simply to the conservation of local structural propensities. The origin of such deviations might arise from several factors, such as protein function or stability, but it is also consistent with the conservation of the type of global non-native attractions suggested by our simulation to be critical to the formation of the trefoil knot. We perform a similar analysis for another parameter, namely the β-sheet propensity (34). As can be seen from Fig. 7*b*, the largest deviation between the SpoU alignment and the reference alignment is again in the middle of helix $\alpha_5$, around positions 139-142 (note that in this propensity scale (34), low values mean a higher propensity for β-sheet structure). It should be pointed out that the hydrophobicity and β-sheet propensity scales are not uncorrelated; strongly hydrophobic residues tend to favor β-sheet structure. Nonetheless, it is interesting that in this region (139-142), which is natively α-helical, the propensity for β-sheet structure is comparable to that of the natively β–sheet structured regions, β4, β5, and β6.

The combination of results from the bioinformatics analysis performed here indicates the following: (*i*) the possibility of nonnative attractive interactions between $\alpha_5$ and the $\beta_5$ regions; and (*ii*) this attraction might transform the natively helical segment 139-142 into a transient $\beta$-sheet structure with $\beta_5$ during the threading of the C-terminal. We point out, however, that hypothesis (*ii*) in particular must be tested rigorously by experiment. It is interesting though, that $\alpha$- to $\beta$-structure transitions have been found in simulations to be expediated by a close spatial proximity of a polypeptide chain in a pre-formed $\beta$-sheet conformation (35). Moreover, amide NH and CO groups of the protein backbone rarely exist in the hydrophobic core of proteins without participating in hydrogen bonds (36), and so the energetics of polypeptide chains may favor the formation of backbone-backbone hydrogen bonds between the "loop" and the "entering" chain of a tight trefoil knot. Indeed, in the native structure of YibK the trefoil knot region is flanked by two $\beta$-sheets (see $\beta_4$ and $\beta_6$ in Fig. 1), and the same is true for the other known structures of SpoU methyltranferases with deep trefoil knots in their C-terminal. Hence, it could be valuable for the C-terminal region, including helix $\alpha_5$, to be able to transiently form $\beta$-sheet structure as the knotted structure is formed. Figure 7 (see Supporting Information) shows more closely the amino acid composition of the $\alpha_5$ helix region, 133-147, in the SpoU alignment. The two most prevalent amino acid types are Alanine and Valine which, in fact, differ sharply in their secondary structure propensities; Alanine strongly favors $\alpha$-helix structure and Valine is frequently observed in $\beta$-sheet structure, which appears consistent with an amino acid segment with an ability to make both $\alpha$- and $\beta$-type secondary structure.

**Conclusion**

We have investigated the folding behavior of the deeply knotted YibK protein through simulations of a coarse-grained $C_\alpha$ model, and by a bioinformatics study of related homologous proteins. Taken together, our results suggest that the folding of YibK depends critically on *specific* attractive, nonnative interactions. Although a complete thermodynamic characterization of the folding process is computationally beyond the scope of our study, a picture of the folding kinetics have been obtained through a large number of folding trajectories, which fits remarkably well with previously obtained experimental data on YibK. As proposed by Mallam and Jackson (11), we see two parallel folding pathways in the first step of the folding process. We attribute these different pathways to an early and a late formation of the trefoil knot. Moreover, we find that the corresponding intermediate states are structurally different, with the "knot-late" pathway involving a more native-like intermediates than the "knot-early" pathway. In order for successful dimerization to occur, we find that knots in the individual monomers have to be formed.

The origin of these critical nonnative interactions may be two conserved hydrophobic regions, in the middle of the trefoil knot ($\beta_5$) and at C-terminal end of the protein ($\alpha_5$). Hydrophobic attractions between these regions fit reasonably well with the optimal energetics of the nonnative $C_\alpha$ model. From our folding simulations, we see that the role of the nonnative interactions is mainly to enhance the kinetic accessibility of the topologically intricate native state of YibK; in the absense of these attractions, the probability of knot formation is exceedingly low. Despite this dramatic impact on folding kinetics, their energetic contributions are relatively small and they likely do not affect strongly the overall thermodynamics. An interesting possibility suggested by the bioinformatics analysis is that an unusually high β-sheet propensity in the $\alpha_5$ helix may prompt the formation of transient β-sheet structure in the C-terminal region, as this chain segment is threaded to form the trefoil knot. Rigorous experimental tests are needed to examine this hypothesis. It would, in particular, be interesting to see if suitable amino acid mutations in positions 129-132 in helix $\alpha_5$ prohibit, or significantly slow down, the folding of the YibK protein.

**Models and Methods**

**Native-centric $C_\alpha$ model**

This $C_\alpha$ chain model was introduced by Clementi et al in 2000 (13). Here we apply it to the residues 1-147 of the YibK protein, as the last 13 residues are unstructured in the native state. The energy function can be expressed as:

$$E_0 = k_{bond} \sum_i (b_i - b_i^0)^2 + k_{bend} \sum_i (\theta_i - \theta_i^0)^2 + \\ \sum_i k_{tors}^{(1)} [1 - \cos(\phi_i - \phi_i^0)] + k_{tors}^{(3)} [1 - \cos 3(\phi_i - \phi_i^0)] + \\ k_{cont} \sum_{ij \in C} \left[ 5 \left( \frac{r_{ij}^0}{r_{ij}} \right)^{12} - 6 \left( \frac{r_{ij}^0}{r_{ij}} \right)^{10} \right] + k_{ev} \sum_{i<j-3} \left( \frac{\sigma_{ev}}{r_{ij}} \right)^{12} \quad (1)$$

where $b_i$, $\theta_i$, and $\phi_i$ denotes the (pseudo) bond lengths, bond angles and torsion angles, respectively, of the $C_\alpha$ chain, and $r_{ij}$ denotes the distance between amino acids i and j. Superscript "0" indicates the corresponding values in the native structure (PDB id: 1J85). The strengths of the various terms are chosen as in previous studies (13, 19), i.e. $k_{bond} = 100\varepsilon$, $k_{bend} = 20\varepsilon$, $k_{tors}^{(1)} = \varepsilon$, and $k_{tors}^{(3)} = 0.5\varepsilon$, and $k_{cont} = k_{ev} = \varepsilon$; $\varepsilon$ sets the energy scale of the model and here we use $\varepsilon = 1.0$. C is a set of native contacts, where amino acids ij are considered in contact if any two of its non-H atoms are within 4.5 Å and $| i - j < 3 |$. Applied to the YibK structure, this gives 327 and 720 native contacts for the monomer and dimer forms, respectively. In the excluded-volume term, the repulsion between nonnative atom pairs is set to $\sigma_{ev} = 4.0$ Å. Dimerization is simulated with two interacting chains in a cube with side length 200 Å and with periodic boundary conditions.

**Nonnative $C_\alpha$ model**

The native-centric model described by Equation 1 is insufficient for our purposes, as it fails to produce the native trefoil knot in YibK (see above). Therefore, we modify its energetics through the addition of attractive, nonnative interactions, i.e. $E = E_0 + E_{nonnat}$, where $E$ is the new energy function. More precisely, we choose

$$E_{nonnat} = -k_{nonnat} \sum_{ij \in \tilde{C}} e^{-(r_{ij}-\sigma_{ev})^2/2} \qquad (2)$$

where $k_{nn} = 0.8\varepsilon$ is the nonnative interaction strength, slightly weaker than the native interactions. The sum goes over a "nonnative contact set" $\tilde{C}$, i.e. a set of amino acid pairs ij which are not in native contact.

We find that this nonnative contact set has to be rather carefully selected in order to obtain a reliable folding behavior for YibK. In this work, we use $\tilde{C} = \{i = 86,...,93; j = 122,...,147\}$. This particular choice of $\tilde{C}$ was found to produce an optimal folding "success rate" among several alternative ways of adding nonnative interactions. An extended discussion of this selection procedure is provided in Supplementary Information.

**Langevin dynamics**

Folding and unfolding kinetics are studied using Langevin dynamics. The means that, for a coordinate $x$, each $C_\alpha$ atom in our model is described by

$$m\dot{v}(t) = -\frac{\partial E}{\partial x} - m\gamma v(t) + R(t) \qquad (3)$$

where $m$, $v(t)$, $-\frac{\partial E}{\partial x}$, $\gamma$, and $R(t)$ are the mass, velocity, conformational force, friction coefficient, and random force, respectively. The random forces $R$ are uncorrelated and drawn from a Gaussian distribution; the variance of this distribution effectively controls the temperature $T$. To numerically integrate Equation 3, we use the velocity form of the Verlet algorithm (see Ref. (37)). All parameters, including friction coefficient $\gamma$, mass $m$, and time-step constant in the numerical integration, are the same as in Refs. (19, 22).

**Bioinformatics**

The approximately 1500 sequences in the SpoU family of MTases were retrieved from the UniProt database (33), and a multiple sequence alignment was built using the ClustalW program with default parameters (32). We compare this SpoU sequence alignment with a reference structural alignment constructed in the following way: Each 9-mer structural segment of YibK is matched against all 9-mer segments in a data set of 1621 nonhomologous protein structures. The middle (i.e. the 5th) amino acid type of the matched segment is included in the alignment if (*i*) the $C_\alpha$ cRMSD between the two segments less than 2 Å and (*ii*) $|K_i - K_j| < 3$, where $K_i$ is the number of intramolecular contacts for the middle $C_\alpha$ atom of segment i ($C_\alpha$-$C_\alpha$ distance less than 8.5 Å). Each matched segment thus adds one amino acid letter to the reference alignment. Criterion (*i*) assures that this amino acid type comes from a segment which is locally highly similar to the corresponding segment in the YibK structure. The second criterion (*ii*) ensures that this segment also is has a local environment, in terms of the number of amino acids in close spatial proximity, which is similar to the situation in YibK.

**Acknowledgments**

We thank Jason E. Donald for help with the standard bioinformatics tools and databases. This work is supported by NIH.

**Figure Captions**

**Figure 1.** Structure of the YibK monomer (PDB id: 1J85). Residues 122-147 (orange) are threaded through a trefoil knot consisting of residues 75-121. In our optimal "nonnative" $C_\alpha$ model, knot formation is mediated by nonnative attraction between residues 122-147 (orange) and 86-93 (magenta). We use the notation of secondary structure elements (e.g. $\alpha_5$) of Ref. (11).

**Figure 2**. Kinetic behavior of the YibK monomer as obtained for the original Gō-type model (13). Two different trajectories are shown: a "folding" run started from an extended, unstructured conformation at $T = 0.975 < T_m$ (a) and an unfolding run started from the native structure at $T = 1.04 > T_m$, where $T_m$ is the melting temperature. $Q$ is the fraction of native contacts ('+'), where two amino acids i and j contact each other if $r_{ij} < 1.2\, r_{ij}^0$, where $r_{ij}$ is the $C_\alpha$-$C_\alpha$ distance between i and j and $r_{ij}^0$ the corresponding native value. The presence of a knot is determined at each MD step, by using KMT reduction (8), and is indicated in the top panels ('□'). In the unfolding trajectory (b), the knot from the native structure disappears after approximately $60 \times 10^6$ MD steps.

**Figure 3**. A typical folding trajectory in $Q$ (a) obtained for the optimal nonnative $C_\alpha$ model, at $T = 0.96$. Also shown is the contribution of native and nonnative contact energy terms (b) during folding (see Models and Methods).

**Figure 4.** Snap shots of a folding event obtained at $T = 0.97$. The time $t$ is given in units of $10^3$ MD steps. The trefoil knot is formed at $t \approx 300$.

**Figure 5**. Probability distribution of the fraction of native contacts in the knot transition state, $Q^*$, at $T = 0.96$. Also shown (inset) is the distributions of the first ($l^*_{first}$) and last ($l^*_{last}$) amino acids of the knots in this transition state.

**Figure 6.** Example of a successful folding and binding trajectory for two interacting YibK chains. The fraction of native contacts, $Q$, for each individual chain is shown for chains 1 (red) and 2 (green). The fraction of inter-chain native contacts is also shown (blue).

**Figure 7.** Profiles of the hydrophobicity (a) and β-sheet propensity (b) obtained for a multiple sequence alignment of the SpoU family (red) and a reference structural alignment (green, see text); booth curves have been weakly smoothed using a sliding window of 3 amino acid positions, allowing

trends along the sequences to be easily seen.

# Figures

**Figure 1**

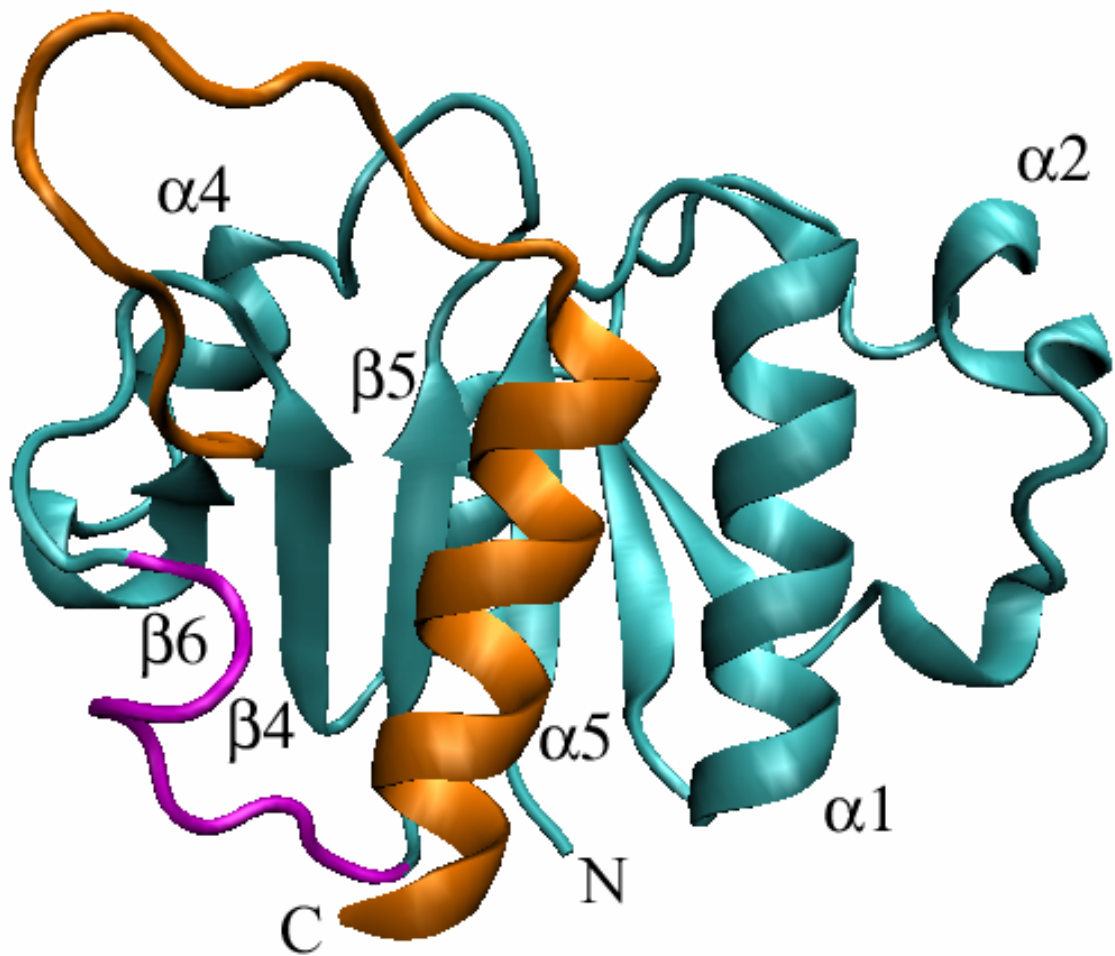

**Figure 2**

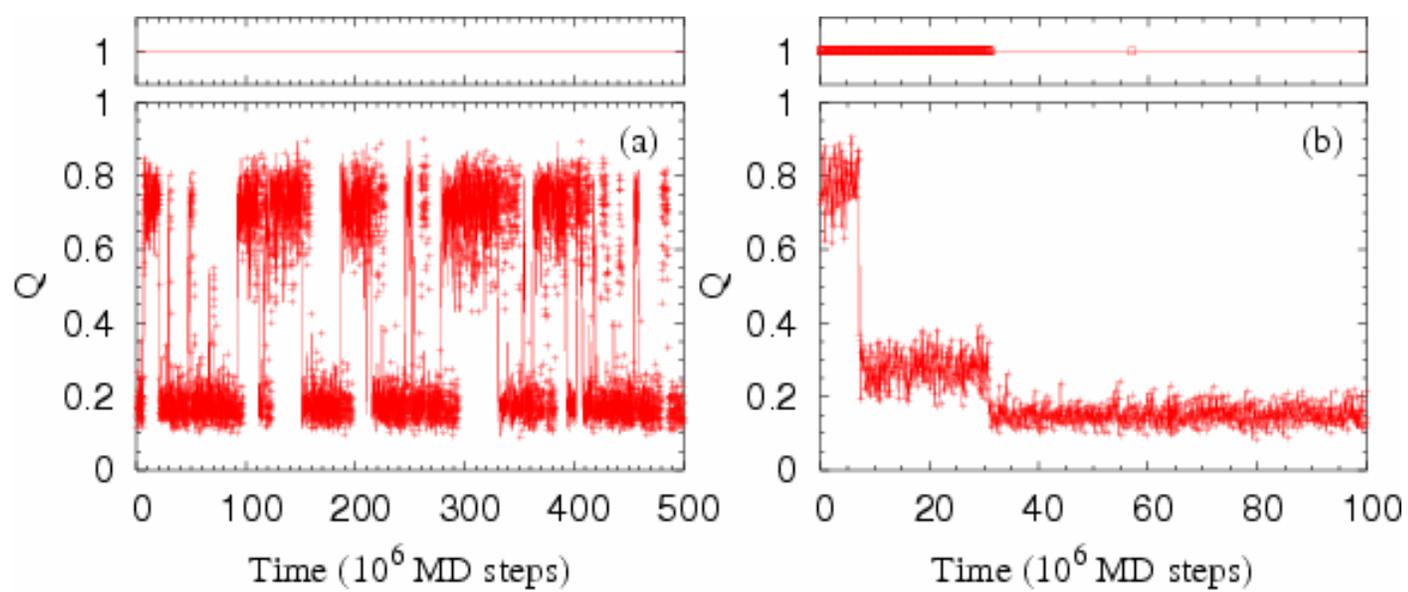

**Figure 3**

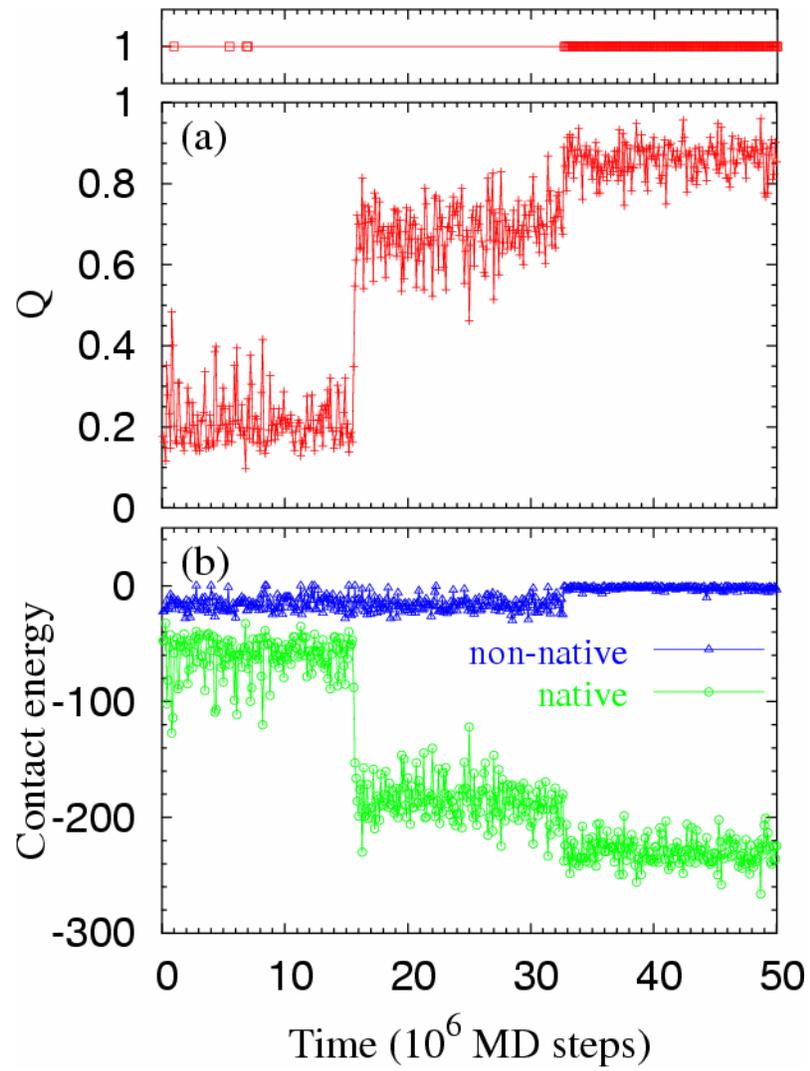

**Figure 4**

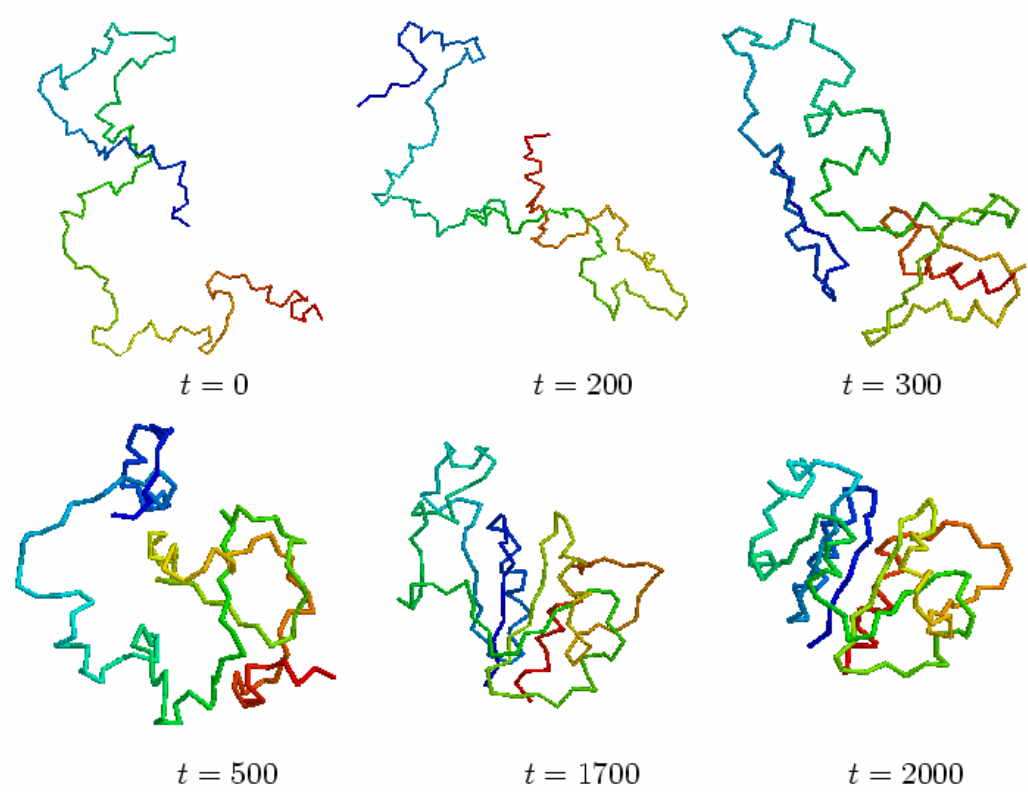

**Figure 5**

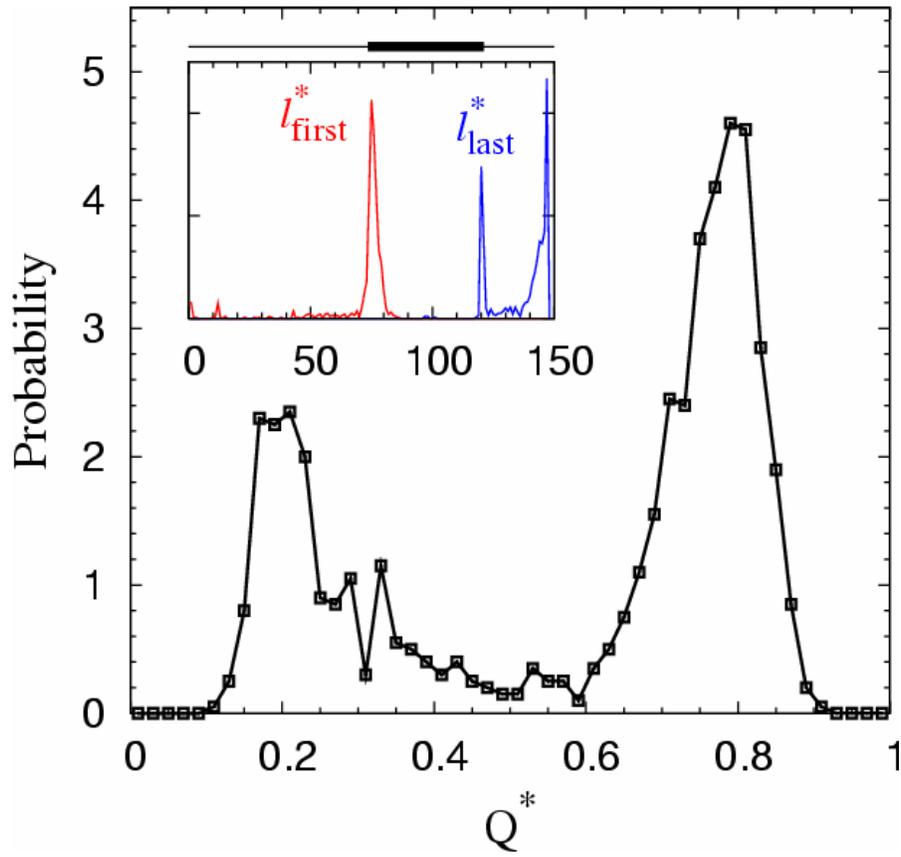

**Figure 6**

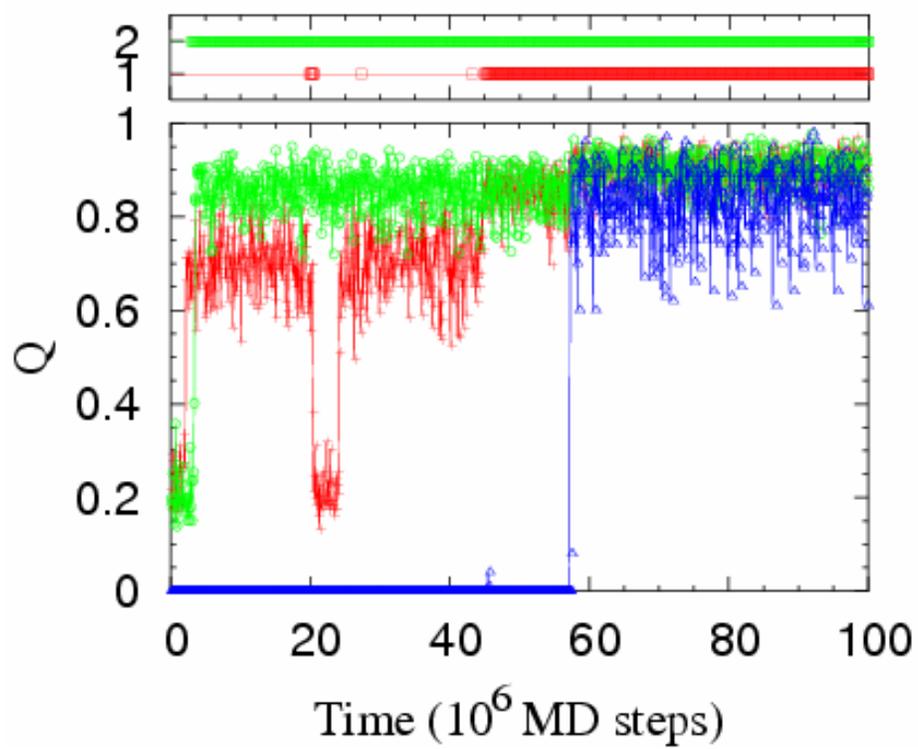

**Figure 7**

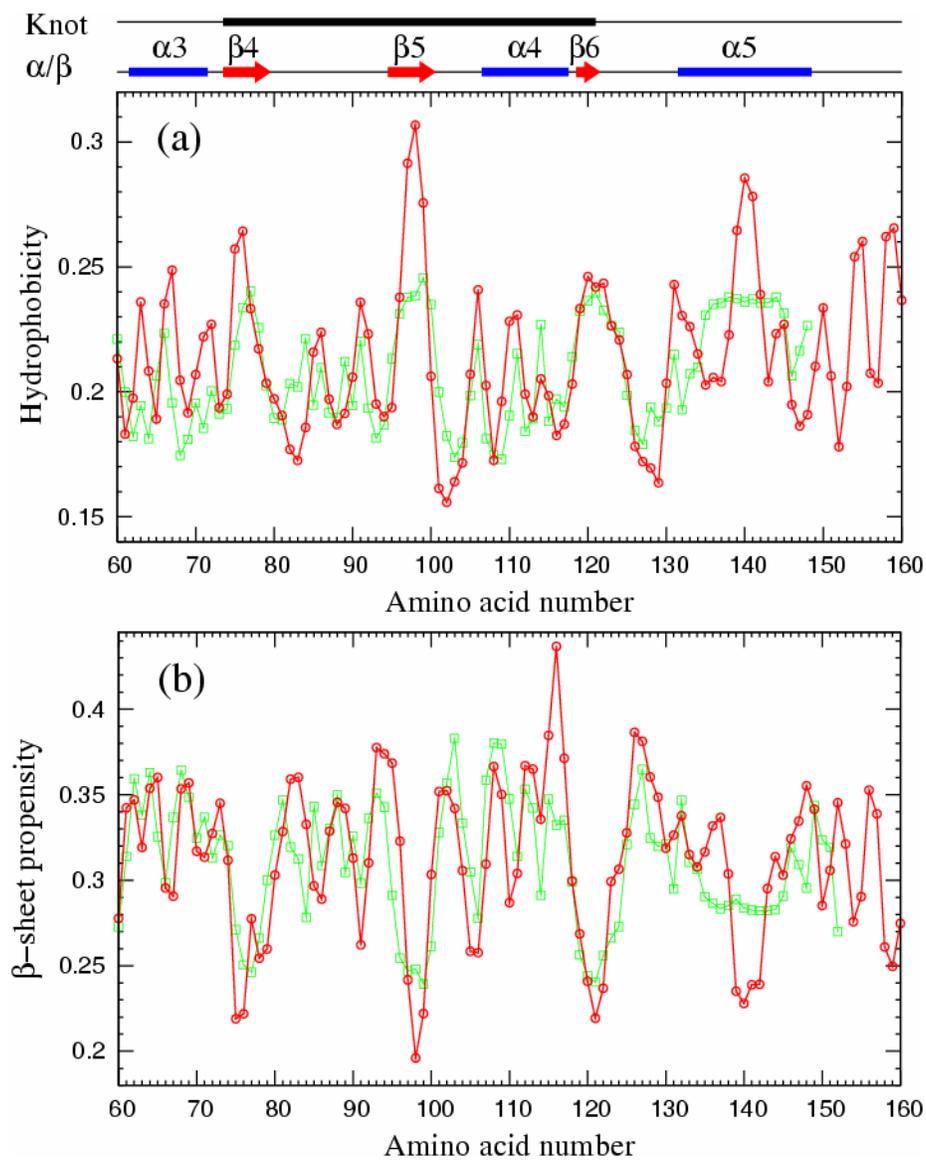

**Supporting Information**

*Effect of nonnative interaction on a native-centric $C_\alpha$ model*

Here we discuss the effect of adding nonnative interactions in various ways to a native-centric $C_\alpha$ model (see Models and Methods). In particular, we investigate the effect of nonnative interactions on the folding behavior of the YibK monomer protein, which contains a deep trefoil knot in its native structure. In this investigation, we quantify the folding "success rate" for a particular model by finding the fraction of 100 independent trajectories started from random, extended conformations at $T = 0.96$, slightly below the melting temperature, that reach the native state N within $10^8$ MD steps; the native state is considered reached when (*i*) $Q > 0.90$, where $Q$ is the fraction of native contacts, and (*ii*) a knot is present according to the KMT reduction procedure (38). For the original native-centric $C_\alpha$ model, none of the performed 100 trajectories reach N. Obviously, the $C_\alpha$ model in its original form is not adequate to study the folding behavior of the YibK protein as no knot is produced (see also Results and Discussion). A model consistent with experiments protein should, in fact, exhibit highly efficiently folding without getting trapped in any intermediate state, as kinetic and thermodynamic experimental data (39, 40) suggest there are no off-pathway intermediates in the folding of YibK.

To test the effect of adding attractive nonnative interactions in different ways to the native-centric $C_\alpha$ model, we construct an energy function $E = E_0 + E_{\text{nonnat}}$, where $E_0$ is the energy function of the original native-centric model and $E_{\text{nonnat}}$ represents the added attractive nonnative interactions, which are controlled by a "nonnative contact set" $\tilde{C}$ (see Models and Methods). We start by testing a set $\tilde{C}$ which includes all nonnative hydrophobic amino acid pairs (Ala, Val, Leu, Ile, Met, Tyr, and Trp are considered hydrophobic). As the hydrophobic effect is one of the main driving forces for folding, it is reasonable to expect that many nonnative hydrophobic contacts are formed during folding, which may be important for knot formation. In an entirely native-centric model, such as the original $C_\alpha$ model, a large number of such hydrophobic attractions are ignored. By re-introducing all hydrophobic attractions in a crude way through the $\tilde{C}$ described above, we find that 15 out of the 100 trajectories in our test procedure reach the native state. This is indeed an improvement from the original model but can not be considered consistent with experimental data, as the remaining 85 trajectories are trapped in off-pathway intermediates. The effect of nonnative interactions on the $C_\alpha$ model has been studied previously by Clementi and Plotkin (41). In Ref. (41), the authors added random, *nonspecific* nonnative interactions using a form of $E_{\text{nonnat}}$ controlled by two parameters $b$ and $\varepsilon_{NN}$, and studied the effect on folding rates. As this nonnative interaction "noise" was found to increase the folding rate for a single-domain protein (at moderate interaction strengths), we apply this procedure to YibK. We use the parameter values $b = 1.0$ and $\varepsilon_{NN} = 0$, which yielded close to optimal increase in folding rates in Ref. (41). No improvement on the success rate for folding for our protein occurs, however, in comparison with the original $C_\alpha$ model. This results, together with the poor performance the "hydrophobic" model, suggests that the addition of more *specific* nonnative forces is needed in order to achieve effective knot formation for the YibK protein.

The trefoil knot in the native structure of YibK is located 40 residues from the C-terminus and 70 residues from the N-terminus. We assume therefore that knot formation occurs at the C-terminal end of the protein. Under this assumption, the C-terminus of the protein chain must enter a loop created by a preceding chain segment during the knot formation part of the folding process, and interactions which mediate such a transition may be critical to folding. Based on this reasoning, we construct a series of models where two separate regions attract each other, namely the region on the C-terminal side of the knot (positions 122-147), which is threaded through a loop during folding, and different proceeding chain segments. Indeed, by adding nonnative attractions this way, we can obtain reliable folding and knot formation for the YibK monomer, as can be seen from Table 1. In particular, we find that attractions involving chain segments around positions 86-93 are very effective at mediating knot formation, a region found roughly in the middle of the native trefoil knot (the native knot stretches positions 75-121). This model, which is then described by $\tilde{C} = \{i = 86...93;\ j = 122...147\}$, is the model used throughout the present work, as it effectively folds our protein without off-pathway intermediate states. Interestingly, nonnative attractions involving regions before the trefoil knot, or in the last part of the knot, produces significantly less effective behavior in terms of reproducing the folding and knotting of YibK (see Table 1).

# Table 1

| Nonnative interactions | Success rate |
|---|---|
| $\tilde{C} = \{i = 54...59, j = 122...147\}$ | 0/100 |
| $\tilde{C} = \{i = 62...69, j = 122...147\}$ | 4/100 |
| $\tilde{C} = \{i = 74...81, j = 122...147\}$ | 62/100 |
| $\tilde{C} = \{i = 86...93, j = 122...147\}$ | 100/100 |
| $\tilde{C} = \{i = 95...102, j = 122...147\}$ | 17/100 |
| $\tilde{C} = \{i = 114...121, j = 122...147\}$ | 3/100 |

# Figure 8

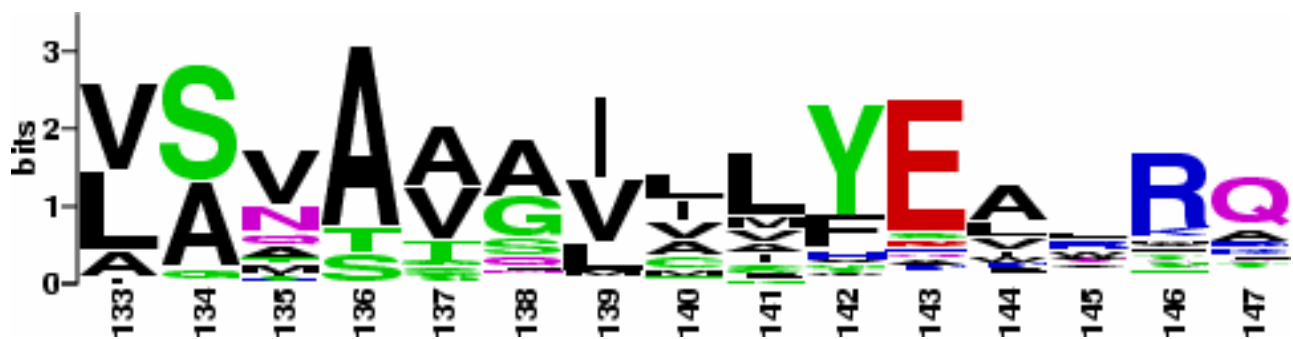

**Figure 8**. Sequence Logo (42) representing the amino acid composition of the $\alpha_5$ helix (residues 133–147) in the SpoU-family of proteins.

30